\documentclass[11pt]{article}
\usepackage{fullpage}
\usepackage{multirow}
\usepackage{amsmath}
\usepackage{amssymb}
\usepackage{amsthm}
\usepackage{mathrsfs}
\usepackage{array}
\usepackage{subfigure}

\usepackage{graphicx}
\usepackage{latexsym,color,amssymb,times}

\usepackage{hyperref}

\newtheorem{theorem}{Theorem}[section]
\newtheorem{lemma}[theorem]{Lemma}
\newtheorem{corollary}[theorem]{Corollary}

\theoremstyle{definition}

\newtheorem{observation}[theorem]{Observation}

\newcommand{\ignore}[1]{}
\newcommand{\remove}[1]{}

\newcommand{\opt}{\text{OPT}}

\usepackage{multirow}
\providecommand{\abs}[1]{\lvert#1\rvert}

\begin{document}

%
% --- Author Metadata here ---
%\conferenceinfo{WOODSTOCK}{'97 El Paso, Texas USA}
%\CopyrightYear{2007} % Allows default copyright year (200X) to be over-ridden - IF NEED BE.
%\crdata{0-12345-67-8/90/01}  % Allows default copyright data (0-89791-88-6/97/05) to be over-ridden - IF NEED BE.
% --- End of Author Metadata ---

\title{Truth and Envy in Capacitated Allocation Games
%\titlenote{With credit to {\sl Fear and Loathing in Las Vegas} by Hunter S. Thompson.}
}
%\subtitle{[Extended Abstract]
%\titlenote{A full version of this paper is available as
%\textit{Author's Guide to Preparing ACM SIG Proceedings Using
%\LaTeX$2_\epsilon$\ and BibTeX} at

\author{
Edith Cohen\thanks{
	AT$\&$T Labs-Research, 180 Park Avenue, Florham Park, NJ.
}
\and
Michal Feldman\thanks{
	School of Business Administration and Center for the Study of Rationality,
The Hebrew University of Jerusalem.
	}
\and
Amos Fiat\thanks{
	The Blavatnik School of Computer Science, Tel Aviv University.
}
\and
Haim Kaplan \thanks{
	The Blavatnik School of Computer Science, Tel Aviv University.}
\and Svetlana Olonetsky\thanks{
The Blavatnik School of Computer Science, Tel Aviv University.
}
}
\date{}
\maketitle 

\begin{abstract}

% Motivated by the problem of paper assignment to program committee
% members,
We study auctions with additive valuations where agents have a limit
on the number of items they may receive.  We refer to this setting
as {\em capacitated allocation games.}  We seek truthful and envy free mechanisms that
maximize the social welfare. {\sl I.e.}, where agents have no incentive to lie and no agent seeks to exchange
outcomes with another.

In 1983, Leonard showed that VCG with Clarke Pivot payments (which is known to be truthful, individually rational, and have no positive transfers), is also an envy free mechanism for the special case of $n$ items and $n$ unit capacity agents. We elaborate upon this problem and show that VCG with Clarke Pivot payments is envy free if agent capacities are all equal. When agent capacities are not identical, we show that there is no truthful and envy free mechanism that maximizes social welfare if one disallows positive transfers.

For the case of two agents (and arbitrary capacities) we show a VCG mechanism that is truthful, envy free, and individually rational, but has
positive transfers. We conclude with a host of open problems that arise from our work.

\end{abstract}

%%%%%%%%%%%%%%%%%%%%%%%%%%%%%%%%%%%%%%%%%%%%%%%%%%%%%%%%%%%%%%
\section{Introduction}

We consider {\em allocation problems} where  a set of objects is to be allocated amongst $m$ agents, where every agent has an additive and non negative valuation function. We study mechanisms that are truthful, envy free, and maximize the social welfare (sum of valuations). The utility of an agent $i$ is the valuation of the bundle assigned to $i$, $v_i(\opt)$, minus any payment, $p_i$.

 A mechanism is incentive compatible (or truthful) if it is a dominant strategy for every agent to report her private information truthfully~\cite{Hurwicz}. A mechanism is envy-free if no agent wishes to switch her outcome with that of another~\cite{dubins,Foley,Svensson,Maskin87,Moulin04,Young95}.

  Any allocation that maximizes the social welfare has payments that make it truthful --- in particular --- any payment of the form
  \begin{equation} p_i = h_i(t^{-i}) - \sum_{j\neq i} v_j(\opt)\label{eq:vcg}\end{equation} where $\opt$ is an allocation maximizing the social welfare and $t^{-i}$ are the types of all agents but agent $i$.  Similarly, any allocation that maximizes the social welfare has payments that make it envy free, this follows from a characterization of envy free allocations (see \cite{Haake2002}). Unfortunately, the set of payments that make the mechanism truthful, and the set of payments that make the mechanism envy free, need not intersect. In this paper we seek such payments, {\sl i.e.}, payments that make the mechanism simultaneously truthful and envy free.

  An example of a mechanism that is simultaneously truthful and envy free is the Vickrey 2nd price auction. Applying the 2nd price auction to an allocation problem assigns items successively, every item going to the agent with the highest valuation to the item at a price equal to the 2nd highest valuation. If, for example, for all items, agent $i$ has maximal valuation, then agent $i$ will receive all items.

  Leonard \cite{Leonard} considered the problem of assigning people to jobs, $n$ people to $n$ positions, and called this problem the permutation game. The Vickrey 2nd price auction is irrelevant in this setting because no person can be assigned to more than one position. Leonard showed that VCG with Clarke Pivot payments is simultaneously truthful and envy free. Under Clarke Pivot payments, agents internalize their externalities, {\sl i.e.}, \begin{equation}h_i(t^{-i}) = \sum_{j\neq i} v_j(\opt^{-i})\label{eq:cpp}\end{equation} where $\opt^{-i}$ is the optimal allocation if there was no agent $i$. By substituting $\sum_{j\neq i} v_j(\opt^{-i})$ for $h_i(t^{-i})$ in Equation \ref{eq:vcg} one can interpret Clarke Pivot payments as though an agent pays for how much others lose by her presence, {\sl i.e.}, the agent internalizes her externalities.

  Motivated by the permutation game, we consider a more general capacitated allocation problem where agents have associated capacities. Agent $i$ has capacity $U_i$ and cannot be assigned more than $U_i$ items. Like Leonard, we seek a mechanism that is simultaneously truthful and envy free.  The private types we consider may include both the valuation and the capacity (private valuations and private capacities) or only the valuation (private valuations, public capacity). Leonard's proof uses LP duality and it is not obvious how to extend it to more general settings.

  Before we address this question, one needs to ask what does it mean for one agent to envy another when they have different capacities? A lower capacity agent may be unable to switch allocations with a higher capacity agent. To deal with this issue, we allow agent $i$, with capacity less than that of agent $i'$ to choose whatever items she desires from the $i'$ bundle, up to her capacity. {\sl I.e.},  we say that agent $i$ envies agent $i'$ if agent $i$ prefers a subset of the allocation to agent $i'$, along with the price set for agent $i'$, over her own allocation and price.

The VCG mechanism (obey Equation \ref{eq:vcg}) is always truthful.
In fact, any
truthful mechanisms that choose the socially optimal allocation in
capacitated allocation problems must be VCG  \cite{Noam07}.
We obtain the following:

  \begin{enumerate}
  \item For agents with private valuations and either private or public capacities, under the VCG mechanism with Clarke Pivot payments, a higher capacity agent will never envy a lower capacity agent. In particular, if all capacities are equal then the mechanism is envy free. (See Section \ref{sec:clarke}).
  \item For agents with private valuations, and either private or public capacities, any envy free VCG payment must allow positive transfers. (See Section \ref{sec:noICEF}).
  \item For two agents with private valuations and arbitrary public capacities, there exist VCG payments such that the mechanism is envy free. It follows that such payments must allow positive transfers. (See Section \ref{sec:ICEFassignment}).
  \item For two agents with private valuations and private capacities, and for two items, there exist VCG payments such that the mechanism is envy free. (See Section \ref{subadditive22:sec}).
  \end{enumerate}

\section{Preliminaries} \label{sec:def}

Let $U$ be a set of objects, and
let $v_i$ be a valuation function
 associated with agent $i$, $1\leq i
\leq m$,  that maps sets of
objects into
 $\Re$.
We denote by $v$ a sequence $<v_1, v_2, \ldots, v_m>$
of valuation
functions one for each agent.

An allocation function\footnote{Here we deal with indivisible allocations,
although our results also extend to divisible allocations with appropriate modifications.} $a$ maps a sequence of valuation functions
$v=<v_1, v_2, \ldots, v_m>$ into a partition of $U$ consisting of
$m$ parts, one for each agent. {\sl I.e.}, $$a(v) = <a_1(v), a_2(v),
\ldots, a_m(v)>,$$ where $\cup_i a_i(v) \subseteq U$ and $a_i(v) \cap a_j(v)
= \emptyset$ for $i \neq j$.
A payment function\footnote{In this
paper we consider only
  deterministic mechanisms and can therefore omit the allocation as an
  argument to the payment function.} is a mapping from $v$ to $\Re^m$,
$p(v) = <p_1(v), p_2(v), \ldots, p_m(v)>$, $p_i(v)\in \Re$. We assume that
payments are from the agent to the mechanism (if the payment is negative then this means that the transfer is from the mechanism to the agent).

A mechanism is a pair of functions, $M=\langle a,p \rangle$, where $a$ is an allocation function, and $p$ is a payment function. For
a sequence of valuation functions $v=\langle v_1, v_2, \ldots, v_m \rangle$, the utility to agent $i$ is defined as $v_i(a_i(v)) - p_i(v)$.
Such a utility function is known as quasi-linear.

Let $v=<v_1, v_2, \ldots, v_m>$ be a sequence of valuations,  we define  $(v'_i,v^{-i})$
to be the sequence of
valuation functions arrived by substituting $v_i$ by $v'_i$, i.e.,
$$(v'_i,v^{-i}) = <v_1, \ldots, v_{i-1}, v'_i, v_{i+1},
\ldots, v_m>.$$

We next define mechanisms that are
incentive compatible, envy-free, and both
incentive compatible and envy-free.

\begin{enumerate}
\item[$\bullet$]
A mechanism  is  {\em incentive compatible\/} ($IC$) if it is a dominant strategy for every agent to reveal her true valuation function to the mechanism. {\sl I.e.}, if for all $i$, $v$, and $v'_i$:
\begin{eqnarray}
v_i(a_i(v)) - p_i(v) \geq v_i(a_i(v'_i,v^{-i})) - p_i(v'_i,v^{-i}); \quad \quad \quad &&\nonumber\\
\Leftrightarrow  p_i(v) \leq p_i(v',v^{-i}) + \Big(v_i(a_i(v)) -
v_i(a_i(v'_i,v^{-i}))\Big). \ &&\label{eq:iccond}
\end{eqnarray}
\item[$\bullet$]
A mechanism  is  {\em envy-free\/} ($EF$) if no agent seeks to switch her allocation and payment with another. {\sl I.e.}, if for all $1\leq i,j\leq m$ and all $v$:
\begin{eqnarray}
&&v_i(a_i(v)) - p_i(v) \geq v_i(a_j(v)) - p_j(v);  \nonumber \\
&&\Leftrightarrow  p_i(v)\leq p_j(v) + \Big(v_i(a_i(v)) -
v_i(a_j(v))\Big). \quad \quad \quad \quad\label{eq:efcond}
\end{eqnarray}
\item[$\bullet$] A mechanism $(a,p)$ is {\sl incentive compatible and envy-free\/} ($IC\cap EF$) if $(a,p)$ is both incentive compatible and envy-free.
\end{enumerate}

\smallskip

\noindent {\bf Vickrey-Clarke-Groves (VCG) mechanism:}
 A mechanism
$M=\langle a,p \rangle$ is called a VCG mechanism if:
\begin{itemize}
\item[$\bullet$] $a(v) \in \mbox{\rm argmax}_{a \in A} \sum_{i=1}^m v_i(a_i(v))$, and
\item[$\bullet$] $p_i(v)= h_i(v^{-i}) - \sum_{j \neq i} v_j(a_j(v))$, where $h_i$ does not depend on $v_i$, $i=1,\ldots,m$.
\end{itemize}

It is known that any mechanism whose allocation function $a$
maximizes $\sum_{i=1}^m v_i(a_i(v))$ (social welfare) is incentive compatible if
and only if it is a VCG mechanism
(See, {\sl e.g.}, \cite{Noam07}, Theorem 9.37).
In the following we will denote by $opt$ an allocation $a$
which maximizes
$\sum_{i=1}^m v_i(a_i(v))$.

The \emph{Clarke-pivot payment}
for a VCG mechanism is defined by $$h_{i}(v^{-i})=\max_{a' \in A} \sum_{j
\neq i} v_j(a').$$

\section{VCG with Clarke-pivot payments} \label{sec:clarke} \label{sec:assignment}

A {\sl capacitated allocation game} has $m$ agents and $n$ items that need to be
assigned to the agents. Agent $i$ is associated with a capacity
$U_i\geq 0$, denoting the limit on the number of items she can be
assigned, and each item $j$ is associated with a capacity $Q_j\geq 0$,
denoting the number of available copies of item $j$. The valuation
$v_i(j)$ denotes how much agent $i$ values item $j$, and $\sum_{j \in
  S}v_i(j)$ is the valuation of agent $i$ to the bundle $S$.

A capacitated allocation
 game has a corresponding bipartite graph $G$, where
every agent $1\leq i\leq m$ has a vertex $i$ associated with it on the
left side, and every item $1\leq j\leq n$ has a vertex $j$ associated
with it on the right side. The weight of the edge $(i,j)$ is
$v_{i}(j)$.  An assignment is a subgraph of $G$ that satisfies the
capacity constraints, i.e.  agent $i$ is assigned at most $U_i$ items
and item $j$ is assigned to at most $Q_j$ agents. Recall that we
denote by $opt$ an assignment of maximum value.  We describe $opt$ by
a matrix $M$ where $M_{ij}$ is the number of copies of item $j$
allocated to agent $i$ in $opt$.

For player $i$, the graph $G^{-i}$ is constructed by removing
the vertex associated with  agent
 $i$ and its incident edges from $G$.  The assignment with maximum
value in $G^{-i}$ is defined by a matrix $M^{-i}$.

Let $M$ be an assignment (either in $G$ or in $G^{-i}$ for some
$i$.). We denote by $M_i$
r the $i$'th row of $M$, $(M_{i1},
M_{i2}, \ldots, M_{in})$
which gives the bundle that agent $i$ gets. We define
$v_k(M_i)=\sum_{j=1}^n M_{ij}v_k(j)$ and $v(M) = \sum_{i=1}^m v_i(M_i)$.

 The Clarke-pivot payment
of agent $k$ is

\begin{equation} \label{cpp:eq}
p_k  =  v(M^{-k}) - v(M) +  v_k(M_{k})\ .
\end{equation}

The main result of this section is that in a VCG mechanism with
Clarke-pivot payments, no agent will ever envy a lower-capacity
agent. In particular, this says that if all agents have the same
capacity, the VCG mechanism with Clarke-pivot payments is both
incentive compatible and envy-free.

The proof of our main result (Theorem \ref{EF:thm}) is
given in terms of a factional assignment but also holds for integral
assignments.

Special case of capacitated allocation games, in which there are $n$ items and $n$
agents, and each agent can get at most a single item was first
introduced in a paper by Leonard~\cite{Leonard}, and was called a {\sl
  permutation game}.  Leonard
proved Theorem \ref{EF:thm} for this special case only, and its proof technique
does not seem to generalize for larger capacities. Our proof is different.

Here is our main theorem.

\begin{theorem}
\label{EF:thm}
Consider a
 VCG mechanism consisting of an optimal allocation $M$ and Clarke-pivot
payments (\ref{cpp:eq}). Then if $U_i\geq U_j$, agent $i$ does not envy
agent $j$.
\end{theorem}

Let agent $1$ and agent $2$ be arbitrary two agents such that the capacity of agent 1 is $\ge$ that of agent 2, that is $U_1 \ge U_2$.

Let $M$ be an optimal assignment,
$M^{-1}$ an optimal assignment without agent $1$,
and $M^{-2}$ some optimal assignment without agent $2$. Agent $1$
does {\sl not} envy agent $2$ iff
\[
v_1(M_{1}) - p_1 \geq v_1(M_{2}) - p_2
\]
Based on Equation~\ref{cpp:eq}, this is true when:
\begin{eqnarray*}
&&v_1(M_{1}) - (v(M^{-1})-v(M)+v_1(M_{1})) = \\
&&v(M) - v(M^{-1})\geq \nonumber \\
&&v_1(M_{2}) -(v(M^{-2}) - v(M) + v_2(M_{2})) = \nonumber \\
&&v_1(M_{2}) + v(M) - v(M^{-2}) - v_2(M_{2})
\end{eqnarray*}
Rearranging we obtain that agent $1$ does not envy agent $2$ iff
\begin{equation}
v(M^{-2}) \ge v(M^{-1})
+ v_1(M_{2}) - v_2(M_{2}). \label{eq:noenvy_gen}
\end{equation}

We prove the theorem by
establishing (\ref{eq:noenvy_gen}). We use the
assignments $M$ and $M^{-1}$ to construct an assignment
$D^{-2}$ on $G^{-2}$ such that

\begin{equation} \label{asswithout2:eq}
v(D^{-2})\geq  v(M^{-1}) + v_1(M_{2})  - v_2(M_{2})\ .
\end{equation}
From the optimality of $M^{-2}$, $v(M^{-2})\geq v(D^{-2})$, which combined with (\ref{asswithout2:eq}) implies
(\ref{eq:noenvy_gen}).

Given assignments $M$ and $M^{-1}$, we construct a flow $f$ on an associated bipartite digraph, $G_f$, with vertices for
every agent and item. We define arcs and flows on arcs in $G_f$ for every agent $i$ and item $j$:
\begin{itemize}
\item If $M_{ij}-M^{-1}_{ij}>0$ then $G_f$ includes an arc $i \rightarrow j$ with flow $f_{i \rightarrow j}=M_{ij}-M^{-1}_{ij}$.
\item If $M_{ij}-M^{-1}_{ij}<0$ then $G_f$ includes an arc $j \rightarrow i$ with flow $f_{j \rightarrow i}=M^{-1}_{ij}-M_{ij}$.
\item If $M_{ij} = M^{-1}_{ij}$ then $G_f$ contains neither $i\rightarrow j$ not $j\rightarrow i$.
\end{itemize}

We define the {\em excess} of an agent $i$ in $G_f$, and the {\em excess} of an item $j$ in $G_f$, to be
\begin{eqnarray*}
ex_i &=& \sum_{(i\rightarrow j)\in G_f} f_{i\rightarrow j} - \sum_{(j\rightarrow i) \in G_f} f_{j\rightarrow i} = \sum_j \big( M_{ij} - M^{-1}_{ij}\big), \\
ex_j &=& \sum_{(j\rightarrow i)\in G_f} f_{j\rightarrow
i}-\sum_{(i\rightarrow j) \in G_f} f_{i\rightarrow j} = \sum_i
\big(M^{-1}_{ij} - M_{ij}\big),
\end{eqnarray*}
respectively.

 In other words the excess is the difference
between the amount flowing out of the vertex  and the amount flowing
into the vertex. Clearly the sum of all excesses is zero. We say
that a node is a {\em source\/} if its excess is positive and we say
that a node is a {\em target\/} if its excess is negative.

\begin{observation} \label{cor:source_target}

To summarize,
\begin{eqnarray}
i \mbox{\ is an agent and a source} &\Rightarrow& \nonumber \\
0 \leq \sum_j M^{-1}_{ij} + \abs{ex_i} &=& \sum_{j} M_{ij} \leq U_i; \label{eq:isource}\\
i \mbox{\ is an agent and a target} &\Rightarrow& \nonumber \\
0 \leq   \sum_j M_{ij} + \abs{ex_i} &=& \sum_j M^{-1}_{ij}  \leq U_i; \label{eq:itarget} \\
j \mbox{\ is an item and a source} &\Rightarrow& \nonumber \\
0 \leq \sum_i M_{ij} + \abs{ex_j} &=& \sum_i M^{-1}_{ij} \leq Q_j; \label{eq:jsource} \\
j \mbox{\ is an item and a target} &\Rightarrow& \nonumber \\
0 \leq  \sum_i  M^{-1}_{ij} + \abs{ex_j}&=& \sum_i M_{ij} \leq Q_j.
\label{eq:jtarget}
\end{eqnarray}

\end{observation}

By the standard flow decomposition theorem we can decompose $f$ into
simple paths and cycles  where each path connects a source to a
target. Each path and cycle $T$  has a positive flow value $f(T)>0$
associated with it. Given an arc $x\rightarrow y$, if we sum the
values $f(T)$ of all paths and cycles $T$ including $x\rightarrow y$ then we
obtain $f_{x\rightarrow y}$.

Notice that $M^{-1}_{1j} = 0$ for all $j$ and therefore $f_{1\rightarrow j}\ge
0$ for all $j$. It follows that there are no arcs of the form
$j\rightarrow 1$ in $G_f$.

\begin{observation} \label{obs:ex}
For each path $P = u_1, u_2, \ldots, u_t$ in flow decomposition $G_f$, where $u_1$ is a source and $u_t$ is a target,
we have $f(P) \leq \min\{ex_{u_1}, \abs{ex_{u_t}}\}$.
\end{observation}

We define the value of a path or a cycle $T = u_1, u_2, \ldots, u_t$ in $G_f$, to be

$$v(P)=\sum_{\begin{array}{c}
        \mbox{agent\ } u_i, \\
        \mbox{item\ } u_{i+1}
      \end{array}} v_{u_i}(u_{i+1}) -
\sum_{\begin{array}{c}
        \mbox{item\ } u_i, \\
        \mbox{agent\ } u_{i+1}
      \end{array}} v_{u_{i+1}}(u_i).$$ It is easy to
verify that the $\sum_T f(T) \cdot v(T)$ over all paths and cycles
in our decomposition is $v(M) - v(M^{-1})$.

\begin{lemma} \label{lem:M^{-1}}Without loss of generality, we can assume that $M^{-1}$ is such that
\begin{enumerate} \item There are no cycles of zero value in $G_f$.
 \item There is no path $P = u_1, u_2, \ldots, u_t$ of zero value such that $u_1\neq 1$ is a source and $u_t$ is a target.
 \end{enumerate}
\end{lemma}
\begin{proof}
Assume that there is a cycle or a path $T$ in the flow decomposition of $G_f$ such that
$v(T)=0$. Let $x$ be the smallest flow along an arc $e$ of $T$. We modify $M^{-1}$ as follows:
For every agent to item arc $i\rightarrow j \in T$ we increase $M^{-1}_{ij}$ by $x$ and for every item to agent arc
$j\rightarrow i \in T$ we decrease $M^{-1}_{ij}$ by $x$. Let the resulting flow be $\tilde{M}^{-1}$.

If $T$ is a cycle then the capacity constraints are clearly
preserved. If $T$ is not a cycle, then the capacity constraints are
trivially preserved for all nodes other than $u_1$ and $u_t$. From
Equation (\ref{eq:isource}) we know that
\begin{eqnarray*}
\sum_j M^{-1}_{u_1j}  &\leq& U_{u_1} - \abs{ex_{u_1}} \leq U_{u_1} - x \mbox{\ if $u_1$ is an agent.}
\end{eqnarray*}
Ergo, if $u_1$ is an agent we can increase the allocation of
$M^{-1}_{u_1u_2}$ by $x$, while not exceeding the capacity of agent $u_1$ ($U_{u_1}$). If $u_1$ is an item, agent $u_2$ can release $x$ units of item $u_1$ without violating any capacity constraints.

We can similarly see that the capacities constraints of $u_t$ are
not violated (Equation (\ref{eq:jtarget})).

Furthermore $v(\tilde{M}^{-1}) = v(M^{-1}) -
x v(T) = v(M^{-1})$ and if we replace $M^{-1}$ by $\tilde{M}^{-1}$ then $G_f$ changes by decreasing the flow along every arc of $T$ by $x$, and removing arcs whose flow becomes zero (in particular at least one arc will be removed). This process does not introduce any new edges to $G_f$.

We repeat the process until $G_f$ does not contain zero cycles or paths as defined.
\end{proof}

From now on we assume that $M^{-1}$ is chosen according to Lemma \ref{lem:M^{-1}}
\footnote{Since Equation (\ref{asswithout2:eq}) depends only on the value of $M^{-1}$ it does not matter which $M^{-1}$ we work with}.
\begin{lemma} \label{lem:nocycle}
The flow $f$ in $G_f$ does not contain cycles.
\end{lemma}
\begin{proof}
Assume that $f$ contains a cycle $C$ which carries $\epsilon > 0$
flow. Clearly $C$ does not contain agent $1$ since there is not any
arc entering agent $1$ in $G_f$.

Assume first that $v(C) < 0$. Create an assignment $\widehat{M}$
from $M$ by decreasing $M_{ij}$ by $\epsilon$ for each agent to item arc
$i\rightarrow j \in C$ and increasing $M_{ij}$ by $\epsilon$ for
each item to agent arc $j\rightarrow i \in C$. This can be done because $M-M^{-1}$
has a flow of $\epsilon$ along the agent to item arc $i\rightarrow j$, so, it must
be that $M_{ij} \geq \epsilon$. Similarly, $M-M^{-1}$ has a flow of $\epsilon$ along item to agent arcs $j \rightarrow i$ so it
must be the $M_{ij} \leq U_i - \epsilon$.
 Since $C$ is a cycle the assignment
$\widehat{M}$ still satisfies the capacity constraints. Furthermore
$v(\widehat{M}) = v(M) -\epsilon v(C) > v(M)$ which contradicts the
maximality of $M$.

If $v(C) > 0$ we create assignment $\widehat{M}^{-1}$ from $M^{-1}$
as follows. For every item to agent arc $j\rightarrow i \in C$ we decrease
$M^{-1}_{ij}$ by $\epsilon$ and for every agent to item  arc $i\rightarrow j \in C$ we increase
$M^{-1}_{ij}$ by $\epsilon$. This can be done because $M^{-1}-M$
has a flow of $\epsilon$ along the item to agent arc $j\rightarrow i$, so, it must
be that $M^{-1}_{ij} \geq \epsilon$. Since $C$ is a cycle $\widehat{M}^{-1}$
still satisfies the capacity constraints. Furthermore
$v(\widehat{M}^{-1}) = v(M^{-1}) + \epsilon v(C) > v(M^{-1})$
which contradicts the maximality of $M^{-1}$.

We need to argue that $\widehat{M}^{-1}$ makes no assignment to agent $1$, this follows
because agent $1$ has no incoming flow in $G_f$ and cannot lie on any cycle.

By assumption, there no cycles of value zero in $G_f$.
\end{proof}

In particular Lemma \ref{lem:nocycle} implies that there are no
cycles in our flow decomposition.

\begin{lemma} \label{lem:1first}
Agent $1$ is the only
source node.% and for every flow path $P$ in $G_f$ from agent $1$ to target $v(P) \geq 0$.
\end{lemma}
\begin{proof}
We give a proof by contradiction, assume some other node, $u_1\neq
1$, is a source. Then, there is a flow path $P= u_1, u_2, \ldots
u_t$ from that node to a target node $u_t$. Since there are no arcs
incoming into vertex $1$, the path $P$ cannot include  agent $1$.

Let $\epsilon$ be the flow along the path $P$ in the flow decomposition.

If $v(P)>0$ define $\widehat{M}^{-1}_{ij} = M^{-1}_{ij}+\epsilon$ for each agent to item arc $i \rightarrow j$  in $P$
and $\widehat{M}^{-1}_{ij} = M^{-1}_{ij}-\epsilon$ for each item to agent arc  $j \rightarrow i$ in $P$. For all other item/agent pairs $(i,j)$, let $\widehat{M}^{-1}_{ij} = M^{-1}_{ij}$. We have that $$v(\widehat{M}^{-1}) = v(M^{-1}) + \epsilon v(P) > v(M^{-1})$$ this would contradict the
maximality of $M^{-1}$ if $\widehat{M}^{-1}$ is a legal assignment.

If $v(P)<0$ define $\widehat{M}_{ij} = M_{ij}-\epsilon$ for each agent to item arc $i \rightarrow j$  in $P$ and $\widehat{M}_{ij} =
M_{ij}+\epsilon$ for each item to agent arc $j \rightarrow i$ in
$P$. For all other item/agent pairs $(i,j)$, let $\widehat{M}_{ij} =
M_{ij}$. We have that
$$v(\widehat{M}) = v(M) - \epsilon v(P) > v(M)$$
which contradicts the maximality of $M$.

We still need to argue that the assignment $\widehat{M}^{-1}$ (if
$v(P)>0$) and the assignment $\widehat{M}$ (if $v(P)<0$) are legal. Because $P$ has a flow of $\epsilon$,
$M^{-1}_{ij} \geq \epsilon$ for each item to agent arc $j\rightarrow i$
along $P$, and $M_{ij} \geq \epsilon$ for each agent to item arc
$i\rightarrow j$ along $P$.

We also worry about exceeding
capacities at the endpoints of $P$, since the size of
assignments of agents/items that are internal to the
path do not change.

We increase the capacity of $u_1$ while constructing ${M}^{-1}$ only if $u_1$ is an agent, and increase the capacity of $u_t$ while constructing ${M}^{-1}$ only if it is an item. By Observation \ref{cor:source_target} this is legal. A similar argument shows that in $\widehat{M}$ the assignment of $u_1$ and $u_t$ is smaller than their capacities.

According to the way we choose $M^{-1}$, it cannot be that $v(P)=0$ and that $P$ carries a flow in $G_f$. \end{proof}

In particular Lemma \ref{lem:1first} implies that all the paths in
our flow decomposition start at agent $1$.

We construct $D^{-2}$ from $M^{-1}$ as follows.
\begin{enumerate}
\item Stage I: Initially, $D^{-2} := M^{-1}$.
\item Stage II:
For each item $j$ let $x=\min\{M_{2j},M^{-1}_{2j}\}$. Set
$D^{-2}_{2j} := M^{-1}_{2j} - x$ and $D^{-2}_{1j} := x$.
\item Stage III:
For each flow path $P$ in the flow decomposition of $G_f$ that contains agent $2$
we consider the prefix of the path up to agent $2$. For each agent to item arc
$i\rightarrow j$  in this prefix we set $D^{-2}_{ij} := D^{-2}_{ij}
+ f(P)$, and for each item to agent arc  $j\rightarrow i$ in this prefix we set
$D^{-2}_{ij} := D^{-2}_{ij} - f(P)$.
\end{enumerate}
It is easy to verify that $D^{-2}$ indeed does not assign any item
to agent $2$. Also, the assignment to agent $1$ in $D^{-2}$ is of the same size as the assignment to agent $2$ in $M^{-1}$. Since $U_1 \geq U_2$, $D^{-2}$ is a legal assignment.

\begin{lemma} \label{lem:satisfy}
The assignment $D^{-2}$ satisfies Equation (\ref{asswithout2:eq}).
\end{lemma}
\begin{proof}
Rearranging Equation (\ref{asswithout2:eq})
\begin{eqnarray}
&& v(D^{-2})\geq v(M^{-1}) \quad \quad \quad \quad \quad \quad  \quad \quad \quad  \quad \quad \quad \quad \quad \quad \quad \label{eq:stage1}\\
&& \ \quad \quad \quad +\sum_{j=1}^n (v_1(j)-v_2(j))\cdot \min(M_{2j},M^{-1}_{2j}) \label{eq:stage2}\\
&& \ \quad \quad \quad +\sum_{j \mid
M_{2j}>M^{-1}_{2j}}(v_1(j)-v_2(j))\cdot
(M_{2j}-M^{-1}_{2j})\label{eq:stage3}.
\end{eqnarray}
At the end of stage I, we have $D^{-2} = M^{-1}$ and so the
inequality above at line (\ref{eq:stage1}) (without adding
(\ref{eq:stage2}) and (\ref{eq:stage3})) holds trivially. It is also easy to verify that at the end
of stage II, the inequality above that spans (\ref{eq:stage1}) and
(\ref{eq:stage2}) but without (\ref{eq:stage3}) holds. Finally, at
the end of stage III, the full inequality in (\ref{eq:stage1}),
(\ref{eq:stage2}) and (\ref{eq:stage3}) will hold as we explain next.

Consider an item $j$ such that $M_{2j} > M^{-1}_{2j}$. In $G_f$ we have an arc $2\rightarrow j$ such
that $f_{2\rightarrow j} = M_{2j} - M^{-1}_{2j}$. Therefore in the flow decomposition we must have
paths $P_1,\ldots, P_\ell$ all containing $2\rightarrow j$ such that
\begin{equation}\sum_{k=1}^\ell f(P_k) = f_{2\rightarrow j} = M_{2j} - M^{-1}_{2j} \label{eq:sumfpk} \end{equation}

Let $\widehat{P}_k$ be the prefix of $P_k$ up to agent $2$. Consider
the cycle $C$ consisting of $\widehat{P}_k$ followed by
$2\rightarrow j$ and $j\rightarrow 1$. It has to be that that value
of this cycle is non-negative. (Otherwise, construct
$\widehat{M}$ by decreasing each agent to item arc $i \rightarrow j$ on the cycle
$\widehat{M}_{ij} = M_{ij}-\epsilon$ and increasing each item to agent arc $j \rightarrow i$ on
the cycle $\widehat{M}_{ij} = M_{ij}+\epsilon$. It follows, that
$v(\widehat{M}) = v(M) - \epsilon v(C) > v(M)$ in contradiction of
maximality of M. The matching $v(\widehat{M})$ is legal since it
preserves capacities and decreases assignment associated with arcs with flow on them.)

Therefore,
\begin{eqnarray*}v(\widehat{P}_k) + v_2(j) -v_1(j) &\geq& 0; \\
\Rightarrow v(\widehat{P}_k) &\geq& (v_{1}(j) - v_2(j)); \\
\Rightarrow f(P_k) v(\widehat{P}_k) &\geq& f(P_k) (v_{1}(j) - v_2(j)); \\
\Rightarrow \sum_{k=1}^\ell \big( f(P_k) v(\widehat{P}_k)\big) &\geq& (v_{1}(j) - v_2(j))\sum_{k=1}^\ell f(P_k).
\end{eqnarray*}
Substituting Equation (\ref{eq:sumfpk}) into the above gives us that
\begin{equation} \sum_{k=1}^\ell \big( f(P_k) v(\widehat{P}_k)\big) \geq \big(v_{1}(j) - v_2(j)\big)\big(M_{2j} - M^{-1}_{2j}\big). \label{eq:sumfpk2} \end{equation}

The left hand side of equation (\ref{eq:sumfpk2}) is exactly the gain in value of the matching when applying stage III to the paths $\widehat{P}_1,\ldots, \widehat{P}_\ell$ during the construction of $D^{-2}$ above. The right hand side is the term which we add in Equation (\ref{eq:stage3}).

To conclude the proof of Lemma \ref{lem:satisfy}, we note that stage
III may also deal with other paths that start at agent 1 and
terminate at agent 2. Such paths must have value $\geq 0$ and thus
can only increase the value of the matching $D^{-2}$. (Otherwise we
can build assignment $\widehat{M}$, such that $v(\widehat{M})
> v(M)$ by decreasing $M_{ij}$ by $\epsilon$ for each arc $i\rightarrow j \in P$ and increasing $M_{ij}$ by
$\epsilon$ for each arc $j\rightarrow i \in P$ as we
did before. The matching $\widehat{M}$ is legal since it preserves
capacities on inner nodes of the path, decreases only arcs with flow
on them, $M_{ij} > \epsilon$. Capacity of a source agent node can be
increased according to Observation \ref{cor:source_target}.)
\end{proof}

\begin{corollary}
\label{coroEF:thm} If all agent capacities are equal then the VCG
allocation with Clarke-pivot payments is envy-free.
\end{corollary}

Do Clarke-pivot payments work also under heterogeneous capacities?
The answer is no.
This follows since in the next section we show that
any mechanism that is both incentive compatible and envy-free
must have positive transfers, and Clarke-pivot payments do not.

%
%
%as demonstrated by the following example. Specifically, it is possible for a lower capacity agent to envy a higher capacity agent. Interestingly, the example shows that envy is possible even when the bundle assigned to the higher capacity agent does not exceed the capacity of the lower-capacity agent.
%
%\begin{example} \label{Michal:ex}
%Consider two agents $1,2$ and two items $1,2$. Agent 1 has valuations $v_{1,1}=1.5$, $v_{1,2}=1$ and capacity $U_1=1$. Agent 2 has valuations $v_{2,1}=1$, $v_{2,2}=2$, and capacity $U_2=2$.  The maximum assignment allocates item $1$ to agent $1$ and item $2$ to agent $2$. The corresponding Clarke Pivot payments are $p_1=1$ and $p_2=0$. Agent $1$ can envy agent $2$, because its utility is $1.5-1=0.5$ but
%is $1-0=1$ on the bundle of agent $2$.
%\end{example}

%TBD

\section{Heterogeneous capacities:  \\ IC$\cap$EF payments imply \\ positive transfers}
\label{sec:noICEF}

Consider an arbitrary VCG mechanism.
Let $$opt=<opt_1, opt_2, \ldots, opt_n>$$ denote the allocation and let
\begin{equation}\label{vcg:eq}
p_i=h_i(v^{-i}) - v^{-i}(opt)
\end{equation}
be the payments, where
$$
\quad v^{-i}(opt) = \sum_{\begin{array}{c}
    1 \leq j \leq n \\
    j \neq i
  \end{array}} v_j(opt_j).$$

Let $v(opt)=\sum_{j=1}^n v_j(opt_j)$
and let
$$opt^{-i} = <opt^{-1}_1, opt^{-1}_2, \ldots, opt^{-i}_{i-1}, \emptyset, opt^{-i}_{i+1}, \ldots, opt_n>,$$ be the allocation maximizing
$$v^{-i}(opt^{-i}) = \sum_{\begin{array}{c}
    1 \leq j \leq n \\
    j \neq i
  \end{array}} v_j(opt^{-i}_j).$$

We
substitute the  VCG payments (\ref{vcg:eq}) into
the envy-free
conditions (\ref{eq:efcond}) and  obtain that
$i$ does not envy $j$ if and only if
\begin{eqnarray}\label{vcgef:eq}
&&v_i(opt_j) - p_j \leq v_i(opt_i) - p_i \nonumber \\
&&\Leftrightarrow p_i - p_j
\leq v_i(opt_i) -v_i(opt_j) \nonumber \\
&&\Leftrightarrow h_i(v^{-i}) - v^{-i}(opt)  - \left(h_j(v^{-j}) - v^{-j}(opt) \right) \nonumber \\
&& \quad \quad \leq v_i(opt_i) - v_i(opt_j)\nonumber \\
&&\Leftrightarrow h_i(v^{-i}) - h_j(v^{-j}) \nonumber \\
&& \quad \quad \leq v^{-i}(opt) - v^{-j}(opt) + v_i(opt_i) - v_i(opt_j)\nonumber \\
&&\Leftrightarrow h_i(v^{-i}) - h_j(v^{-j}) \nonumber \\
&& \quad \quad \leq v(opt) - \left(v(opt) - v_j(opt_j)\right)  - v_i(opt_j)\nonumber \\
&&\Leftrightarrow h_i(v^{-i}) - h_j(v^{-j}) \leq v_j(opt_j) - v_i(opt_j). \label{eq:mainenvyfree}
\end{eqnarray}

  \begin{theorem} \label{thm:pt}
 Consider a
capacitated allocation game with heterogeneous capacities such that
the number of items exceeds the smallest agent capacity.  There is no
mechanism that simultaneously optimizes the social welfare,
is IC$\cap$EF, and has no positive transfers
(the mechanism never pays the agents).  That is, any IC$\cap$EF mechanism
has some valuations $v$ for which the mechanism pays an agent.
  \end{theorem}

 Note that the conditions on the  capacities
of the agents and the number of items
are necessary -- If capacities are homogeneous or the total supply
of items is at most the minimum agent capacity then Clarke-pivot
payments, that are known to be incentive compatible,
individually rational, and
have no positive transfers, are also envy-free.

In the rest of this section we prove Theorem \ref{thm:pt}.
We start with
a capacitated allocation game with two agents and two items
where agent $i$ has capacity $i$ ($i=1,2$).  We then
generalize the proof to arbitrary heterogeneous games.

To ease the notation we abbreviate in the rest of the paper
$v_i(j)$ to $v_{ij}$.

We partition the valuations into three sets
$A$, $B_1$, and $B_2$ as follows (we omit cases with
ties).\footnote{The optimal allocation that maximizes social welfare
  is uniquely defined when there are no ties.  Valuations $v$'s with
  ties form a lower dimensional measure 0 set.  It suffices to
  consider valuations without ties for both existence or non-existence
  claims of $IC$ or $EF$ payments.  This is clear for non-existence, for
  existence, the payments for a $v$ with ties is defined as the limit
  when we approach this point through $v$'s without ties that result
  in the same allocation. Clearly IC and EF properties carry over,
  also IR and nonnegativity of payments.}
\begin{itemize}
\item[$\bullet$](A)
$v_{21}> v_{11}$ and $v_{22}>v_{12}$.
For these valuations in an optimal allocation
agent $2$ obtains the bundle $\{1,2\}$ and agent $1$ obtains the
empty bundle.
\item[$\bullet$](B$_1$)
$v_{11}-v_{21} > \max\{0,v_{12}-v_{22}\}$.
For these valuations in an optimal allocation
item $1$ is assigned to agent $1$ and item $2$ to agent $2$.
\item[$\bullet$](B$_2$)
$v_{12}-v_{22} > \max\{0,v_{11}-v_{21}\}$.
For these valuations in an optimal allocation
item $1$ is assigned to agent $2$ and item $2$ to agent $1$.
\end{itemize}

  Substituting the above in (\ref{eq:mainenvyfree})
we obtain that for $v\in B_1$, agent 1 does not envy agent 2 if
and only if
$$h_1(v_2) - h_2(v_1) \leq v_2(opt_2) - v_1(opt_2) = v_{22}-v_{12}\ .$$
Agent 2 does not envy agent 1 if and only if
$$h_2(v_1) - h_1(v_2) \leq v_1(opt_1) - v_2(opt_1) = v_{11}-v_{21}\ .$$
Combining we obtain that there is no envy for $v\in B_1$, if and only
if
\begin{equation}  \label{B1_noenvy}
v_{21}-v_{11} \leq h_1(v_2) - h_2(v_1) \leq  v_{22}-v_{12}\ .
\end{equation}

  For a fixed $\epsilon>0$, and $x> 5\epsilon$, the valuation $v$ such that
$v_{11}=x+3\epsilon$, $v_{12}=x+\epsilon$, $v_{21}=v_{22}=0$
is clearly in $B_1$.
Substituting in~(\ref{B1_noenvy})  we obtain
\begin{equation}  \label{cc1}
-(x+3\epsilon)  \leq  h_1(0,0)-h_2(x+3\epsilon,x+\epsilon) \leq -(x+\epsilon)
\end{equation}.

 The valuation $v$ such that
$v_{11}=x+3\epsilon$, $v_{12}=x+\epsilon$, $v_{21}=x+\epsilon$, and $v_{22}=x$
is also clearly in $B_1$ and from (\ref{B1_noenvy})  we obtain
\begin{equation*}
x+\epsilon-(x+3\epsilon) \leq h_1(x+\epsilon,x)-h_2(x+3\epsilon,x+\epsilon) \leq  x-(x+\epsilon)
\end{equation*} hence
\begin{equation} \label{cc2}
-2\epsilon \leq h_1(x+\epsilon,x)-h_2(x+3\epsilon,x+\epsilon) \leq -\epsilon\ .
\end{equation}

 Combining (\ref{cc1}) and (\ref{cc2}) we obtain
\begin{equation} \label{cc3}
h_1(x+\epsilon,x) \leq h_2(x+3\epsilon,x+\epsilon) -\epsilon
                  \leq h_1(0,0) + x+3\epsilon
\end{equation}

  The no positive transfers requirement is that for any $v$,
\begin{equation} \label{npt1}
h_1(v_2)  \geq  v_2(opt_2)\ .
\end{equation}
 Consider now the valuations $v$ such that
$v_{21}=x+\epsilon$, $v_{22}=x$, $v_{11}=v_{12}=x-\epsilon$.
Clearly, $v\in A$ (agent $2$ gets both items), hence $v_2(opt_2)=2x-\epsilon$.
Substituting this and (\ref{cc3}) in (\ref{npt1}) we obtain
$2x-\epsilon \leq h_1(0,0) + x+3\epsilon$, hence $h_1(0,0)\geq x-4\epsilon$.
Clearly, for valuations with large enough $x$ we obtain a contradiction, that is, there exist valuations where the mechanism pays an agent.

\noindent
{\bf Heterogeneous capacities, multiple agents and items:}
Let $c$ be the smallest agent capacity and assume it is the capacity
of agent 1.  Let agent $2$ be any agent with capacity $>c$.
There are $\geq c+1$ items.
It suffices to consider restricted valuation matrices $v$
where
$v_{ij}=0$ when $i>2$ or when $j>c+1$  and
$v_{ij}\equiv v_{i2}$ for $i=1,2$ and $2\leq j\leq c+1$.
We partition these valuations into four sets $A$, $B_1$, $B_1^+$, $B_2$, as follows
 (we omit cases with ties and
only define the assignment of items $1,\ldots,c+1$):
\begin{itemize}
\item[$\bullet$](A)
$v_{21}> v_{11}$ and $v_{22}>v_{12}$.
For these valuations in an optimal allocation
agent $2$ obtains the bundle $\{1,\ldots,c+1\}$.
\item[$\bullet$](B$_1$)
$v_{11}>v_{21}$ and $v_{12}<v_{22}$.
For these valuations in an optimal allocation
items $1$ is assigned to agent $1$ and items $2,\ldots,c+1$ to agent $2$.
\item[$\bullet$](B$_1^+$)
$v_{11}-v_{21} > v_{12}-v_{22}$ and $v_{12}>v_{22}$.
For these valuations in an optimal allocation
items $1,\ldots,c$ are assigned to agent $1$ and item $c+1$ is assigned to agent $2$.
\item[$\bullet$](B$_2$)
$v_{12}-v_{22} > \max\{0,v_{11}-v_{21}\}$.
For these valuations in an optimal allocation
item $1$ is assigned to agent $2$ and items $2,\ldots,c+1$ to agent $1$.
\end{itemize}

  Substituting the above in (\ref{eq:mainenvyfree})
we obtain that for $v\in B_1^+$, agent 1 does not envy agent 2 if
and only if
$$h_1(v_2) - h_2(v_1) \leq v_2(opt_2) - v_1(opt_2) = v_{22}-v_{12}\ .$$
Agent 2 does not envy agent 1 if and only if
\begin{eqnarray*}
h_2(v_1) - h_1(v_2) & \leq & v_1(opt_1) - v_2(opt_1) \\
 & = & v_{11}+(c-1)v_{12} -v_{21}-(c-1)v_{22}\ .
\end{eqnarray*}
Combining we obtain that there is no envy for $v\in B^+_1$, if and only
if
\begin{equation}  \label{B1p_noenvy}
v_{21}+(c-1)v_{22}-v_{11}-(c-1)v_{12} \leq h_1(v_2) - h_2(v_1) \leq  v_{22}-v_{12}\ .
\end{equation}

  For a fixed $\epsilon>0$ and for $x>\epsilon$, the valuation $v$ such that
$v_{11}=x+3\epsilon$, $v_{12}=x+\epsilon$, $v_{21}=v_{22}=0$
is clearly in $B_1^+$.
For such $v$ the left hand side of (\ref{B1p_noenvy}) is
\begin{eqnarray*}
\lefteqn{v_{21}+(c-1)v_{22}-v_{11}-(c-1)v_{12} } \\
 & = & -(x+3\epsilon)-(c-1)(x+\epsilon) \\
  & = & -cx-(c+2)\epsilon
\end{eqnarray*}
Substituting in~(\ref{B1p_noenvy})  we obtain
\begin{equation}  \label{cc1g}
-cx-(c+2)\epsilon \leq  h_1(0,0)-h_2(x+3\epsilon,x+\epsilon) \leq -(x+\epsilon)\ .
\end{equation}

 The valuation $v$ such that
$v_{11}=x+3\epsilon$, $v_{12}=x+\epsilon$, $v_{21}=x+\epsilon$, and $v_{22}=x$
is also clearly in $B_1^+$.
For such $v$ the left hand side of (\ref{B1p_noenvy}) is
\begin{eqnarray*}
\lefteqn{v_{21}+(c-1)v_{22}-v_{11}-(c-1)v_{12}}\\
& = &    x+\epsilon+(c-1)x -(x+3\epsilon) -(c-1)(x+\epsilon)\\
  & = & -(c+1)\epsilon
\end{eqnarray*}
From (\ref{B1p_noenvy})  we obtain
\begin{equation} \label{cc11g}
-(c+1)\epsilon  \leq h_1(x+\epsilon,x)-h_2(x+3\epsilon,x+\epsilon) \leq  -\epsilon .
\end{equation}

Combining (\ref{cc1g}) and (\ref{cc11g}) we obtain,
\begin{eqnarray*}
h_1(x+\epsilon,x) & \leq & h_2(x+3\epsilon,x+\epsilon)-\epsilon\\
& \leq & h_1(0,0)+cx+(c+2)\epsilon-\epsilon\\
& = & h_1(0,0)+cx+(c+1)\epsilon
\end{eqnarray*}

For valuations $v_{21}=x+\epsilon$, $v_{22}=x$, $v_{11}=v_{12}=x-\epsilon$,
we clearly have $v\in A$ (agent $2$ gets all items), hence
$v_2(opt_2)=(c+1)x-(c+1)\epsilon$.

 For a sufficiently large $x$ (relative to $\epsilon$ and $h_1(0,0)$),
$h_1(v_2)=h_1(x+\epsilon,x)\leq h_1(0,0)+cx+(c+1)\epsilon < (c+1)x-(c+1)\epsilon = v_2(opt_2)$, which contradicts the
no positive transfers requirement (\ref{npt1}).

\section{2 agents, Public Capacities}
\label{sec:ICEFassignment}

In this section we assume that capacities are public and
derive $IC\cap EF$ payments for any game with two players.

\begin{lemma} \label{2agent:lemma}
 Any 2-player capacitated allocation game with public capacities has an $IC\cap EF$
individually rational mechanism.
\end{lemma}
\begin{proof}
Let $c_i$ be the capacity of player $i$ and assume without loss of generality that  $c_1\leq c_2$.
For a vector $(x_1,x_2\ldots)$ let
$top_{b}\{x\}$ be the set  of the $b$ largest entries in $x$.
We show that
$$h_1(v_2) = \sum_{j\in top_{c_1}\{v_2 \}} v_{2j}$$ and
$$h_2(v_1) = \sum_{j\in top_{c_1}\{v_1 \}} v_{1j}$$ give VCG payments which
are envy-free.

 It suffices to show that for $\{i,j\}=\{1,2\}$,
$$ h_i(v^{-i}) - h_j(v^{-j}) \leq v_j(opt_j) - v_i(opt_j)\ .$$
That is,
\begin{equation}
\sum_{j\in top_{c_1}\{v_2 \}} v_{2j} -\sum_{j\in top_{c_1}\{v_1 \}} v_{1j}  \leq  v_2(opt_2)-v_1(opt_2) \label{env12}
\end{equation}
and
\begin{equation}
\sum_{j\in top_{c_1}\{v_1 \}} v_{1j} -\sum_{j\in top_{c_1}\{v_2 \}} v_{2j} \leq v_1(opt_1)-v_2(opt_1) \label{env21} .
\end{equation}

  Assume first that the number of items is exactly $c_1+c_2$.
In the optimal solution, player $1$ will get the $c_1$ items that
maximize $v_{1j}-v_{2j}$ and player $2$ will get the $c_2$ items
that minimize this difference.

 We establish (\ref{env21}) as follows
\begin{eqnarray*}
\lefteqn{\sum_{j\in top_{c_1}\{v_1 \}} v_{1j} - \sum_{j\in top_{c_1}\{v_2 \}} v_{2j}}\\
 & \leq & \sum_{j\in top_{c_1}\{v_1 \}} (v_{1j}-v_{2j}) \\
& \leq & \sum_{j\in top_{c_1} \{v_1-v_2\}} (v_{1j}-v_{2j}) \\
& = & \sum_{j\in opt_1} (v_{1j}-v_{2j}) = v_1(opt_1)-v_2(opt_1) \ .
\end{eqnarray*}

 We establish (\ref{env12}) as follows
\begin{eqnarray*}
\lefteqn{\sum_{j\in top_{c_1}\{v_2 \}} v_{2j} - \sum_{j\in top_{c_1}\{v_1 \}} v_{1j}}\\
 & \leq & \sum_{j\in top_{c_1}\{v_2 \}} (v_{2j}-v_{1j}) \\
& \leq & \sum_{j\in top_{c_1} \{v_2-v_1\}} (v_{2j}-v_{1j}) \\
& \leq & \sum_{j\in opt_2} v_{2j} - \sum_{j\in top_{c_1} \{v_{1}(opt_2)\}} v_{1j}
\end{eqnarray*}
where $v_1(opt_2)$ is the vector of the values of player $1$ to the items
player 2 gets in the optimal solution.

  If there are fewer than $c_1+c_2$ items, we add ``dummy'' items with
valuations $v_{1j}=v_{2j}=0$ and
the lemma follows from the previous argument for the case with
$c_1+c_2$ items.

  If there are more than $c_1+c_2$ items then
 consider the set of
$c_1+c_2$ items than participate in the optimal solution.
  We now observe that (\ref{env12}) and (\ref{env21})  only involve items
that participate in the optimal solution ($top_{c_2}\{v_2 \}$ and $top_{c_1}\{v_1 \}$ must both be included in
the optimal solution).
\end{proof}

\section{2 agents, 2 items, Private Capacities} \label{subadditive22:sec}

In this section, valuations and capacities are private. We give
 VCG payments which are envy-free and individually rational for
any game with two agents and two items. We
specify the payments by giving the  functions
$h_1(v_2,c_2)$ and $h_2(v_1,c_1)$. Note that
with two items, all $c_i\geq 2$
are equivalent, therefore we only need to consider
capacities $\in\{0,1,2\}$.

% ignore

  We show that the following give  envy-free payments
\begin{eqnarray*}
  h_1(v_2,c_2) &=& \left\{ \begin{array}{ll}
      \max(v_{21},v_{22}) & c_2\in\{1,2\} \\
      0 & c_2=0 \\
    \end{array} \right.\\
  h_2(v_1,c_1) &=& \left\{ \begin{array}{ll}
      \max(v_{11},v_{12}) & c_1\in\{1,2\} \\
      0 & c_1=0 \\
    \end{array} \right.
\end{eqnarray*}

The payments are envy-free if and only if
\begin{eqnarray*}
 \delta_{12} =  h_1(v_2,c_2) - h_2(v_1,c_1)  &\leq& v_2(opt_2) - v_1(opt_2), \\
  \delta_{21} = h_2(v_1,c_1) - h_1(v_2,c_2)  &\leq& v_1(opt_1) - v_2(opt_1).
\end{eqnarray*}

 The conditions when $\{c_1,c_2\}=\{1,2\}$ were worked out
 in the previous section
 and the correctness for $h_1(v_2,2)$ and $h_2(v_1,1)$ carries over (and symmetrically, if
we switch capacities of the agents).
Consider the following remaining cases.

\medskip\noindent
{\bf $\bullet$ $c_1=c_2=2$:}
agent 1 does not envy agent 2 if and only if:
\begin{eqnarray*}
\lefteqn{h_1(v_2,2) - h_2(v_1,2)  \leq} \\
& &  \left\{ \begin{array}{ll}
          v_{21}+v_{22} - v_{11}-v_{12} & \mbox{if } v_{21}>v_{11}, v_{22}>v_{12} \\
          v_{22} - v_{12}                &\mbox{if } v_{21}<v_{11}, v_{22}>v_{12} \\
          v_{21} - v_{11}                &\mbox{if } v_{21}>v_{11}, v_{22}<v_{12} \\
          0                              &\mbox{if } v_{21}<v_{11}, v_{22}<v_{12} \\
        \end{array} \right.
\end{eqnarray*}

Symmetrically, agent 2 does not envy agent 1 if and only if:
\begin{eqnarray*}
\lefteqn{h_2(v_1,2) - h_1(v_2,2)  \leq} \\
& & \left\{ \begin{array}{ll}
          v_{11}+v_{12} - v_{21}-v_{22} & \mbox{if } v_{11}>v_{21}, v_{12}>v_{22} \\
          v_{12} - v_{22}                & \mbox{if } v_{11}<v_{21}, v_{12}>v_{22} \\
          v_{11} - v_{21}                &\mbox{if } v_{11}>v_{21}, v_{12}<v_{22} \\
          0                              &\mbox{if } v_{11}<v_{21}, v_{12}<v_{22} \\
        \end{array} \right.
\end{eqnarray*}

 Combining, we obtain the condition
\begin{eqnarray}
\lefteqn{\min\{v_{21}-v_{11},0\}+\min\{v_{22}-v_{12},0\}} \nonumber \\
 & \leq & h_1(v_2,2)-h_2(v_1,2) \nonumber \\
& \leq & \max\{v_{21}-v_{11},0\}+\max\{v_{22}-v_{12},0\} . \label{22envy:ineq}
\end{eqnarray}

  We now show that our particular $h$'s satisfy (\ref{22envy:ineq}).  It
suffices to establish one of the inequalities:
We have
\begin{eqnarray*}
v_{21} & \leq &\max\{v_{11},v_{12}\} + \max\{v_{21}-v_{11},0\} \\
v_{22} & \leq & \max\{v_{11},v_{12}\} + \max\{v_{22}-v_{12},0\}
\end{eqnarray*}
Combining, we obtain the desired relation:
\begin{eqnarray*}
\lefteqn{\max\{v_{21},v_{22}\}}\\
 & \leq & \max\{v_{11},v_{12}\}+ \max\{v_{21}-v_{11},0\}+\max\{v_{22}-v_{12},0\}\ .
\end{eqnarray*}

\medskip\noindent
{\bf $\bullet$ $c_1=c_2=1$:}
agent 1 does not envy agent 2 if and only if:
\begin{eqnarray*}
\lefteqn{h_1(v_2,1) - h_2(v_1,1)}\\
  &\leq& \left\{ \begin{array}{ll}
          v_{22} - v_{12} & v_{11}+v_{22} > v_{12} + v_{21} \\
          v_{21} - v_{11} & v_{11}+v_{22} < v_{12} + v_{21}
                                                     \end{array} \right.
\end{eqnarray*}
Symmetrically, agent 2 does not envy agent 1 if and only if:
\begin{eqnarray*}
\lefteqn{h_2(v_1,1) - h_1(v_2,1)}\\
  &\leq& \left\{ \begin{array}{ll}
          v_{12} - v_{22} & v_{21}+v_{12} > v_{22} + v_{11} \\
          v_{11} - v_{21} & v_{21}+v_{12} < v_{22} + v_{11}
                                                     \end{array} \right.
\end{eqnarray*}

 Combining, we obtain
\begin{eqnarray}
\lefteqn{\min\{v_{22}-v_{12},v_{21}-v_{11}\}} \nonumber \\
 & \leq & h_1(v_2,1)-h_2(v_1,1) \nonumber \\
& \leq & \max\{v_{22}-v_{12},v_{21}-v_{11}\} \label{11envy:ineq}
\end{eqnarray}

  We now show that our particular $h$'s satisfy (\ref{11envy:ineq}).  It
suffices to establish one of the inequalities:
We have
\begin{eqnarray*}
v_{21} & \leq &\max\{v_{11},v_{12}\} + v_{21}-v_{11} \\
v_{22} & \leq & \max\{v_{11},v_{12}\} + v_{22}-v_{12}\}
\end{eqnarray*}
Combining, we obtain the desired relation:
$$\max\{v_{21},v_{22}\} \leq \max\{v_{11},v_{12}\}+ \max\{v_{21}-v_{11},v_{22}-v_{12}\}\ .$$

\medskip\noindent
{\bf $\bullet$ $c_1=1, c_2=0$:}
No agent envies the other if and only if
\begin{eqnarray*}
h_1(v_2,0)-h_2(v_1,1) & \leq & 0 \\
h_2(v_1,1)-h_1(v_2,0) & \leq & \max\{v_{11},v_{12}\} \\
\end{eqnarray*}

 Combining, we obtain
\begin{equation} \label{10envy:ineq}
-\max\{ v_{11},v_{12} \}\leq h_1(v_2,0)-h_2(v_1,1) \leq 0
\end{equation}

 Symmetrically, when $c_1=0, c_2=1$:
\begin{equation} \label{01envy:ineq}
-\max\{ v_{21},v_{22} \}\leq h_2(v_1,0)-h_1(v_2,1) \leq 0
\end{equation}

Our particular $h$'s trivially satisfy (\ref{10envy:ineq}) and (\ref{01envy:ineq}).

\medskip\noindent
{\bf $\bullet$ $c_1=2, c_2=0$:}
No agent envies the other if and only if
\begin{eqnarray*}
h_1(v_2,0)-h_2(v_1,2) & \leq & 0 \\
h_2(v_1,2)-h_1(v_2,0) & \leq & v_{11}+v_{12} \\
\end{eqnarray*}

 Combining, we obtain
\begin{equation} \label{20envy:ineq}
-v_{11}-v_{12}\leq h_1(v_2,0)-h_2(v_1,2) \leq 0
\end{equation}

 Symmetrically, when $c_1=0, c_2=2$:
\begin{equation} \label{02envy:ineq}
-v_{21}-v_{22} \leq h_2(v_1,0)-h_1(v_2,2) \leq 0
\end{equation}

Our particular $h$'s trivially satisfy (\ref{20envy:ineq}) and (\ref{02envy:ineq}).

\ignore{
{\bf Edith says:

  It seems that these max $h$'s also work for any combinatorial auction with 2 agents and 2 items such that the valuation function are subadditive.  I need to put a proof.  By subadditive I mean that
the value of an agent on getting both items is at most the sum of the
valuations.

  It is not clear what happens with superadditive valuations.  If we can
show that there are no $h$'s, we will have an example of an allocation that
optimizes the social welfare where
there are no VCG payments that are IC$\cap$EF.
Any allocation that optimizes social welfare is IC$\cup$EF.

 Noam gave Michal an example of a 3 player game with superadditive valuations
that is not $IC \cap EF$.  Probably for 2$\times$2 superaditive valuations
is also not $IC \cap EF$.??

}}

\section{Conclusion and open problems} \label{beyond:sec}

We have begun to study truthful and  envy free mechanisms for maximizing
social welfare for the capacitated allocation problem.

There is much left open, for example:
\begin{enumerate}
\item Is there a truthful and envy free mechanism (with positive transfers) for the capacitated allocation problem (arbitrary capacities):
    \begin{enumerate}
    \item  With public capacities and more than two agents.
    \item With private capacities for more than 2 agents and 2 items?
    \end{enumerate}
  \item
How well can we approximate the social welfare by
a mechanism that is incentive-compatible, envy-free, invidually
rational, and without positive transfers for capacitated allocations ?
\item Noam Nisan has observed that for superadditive valuations, there may be no mechanism that is both truthful and envy free. We conjecture that one can obtain mechanisms that are both truthful and envy free for subadditive valuations.
   \end{enumerate}

\bibliographystyle{plain}
\bibliography{envy}

\end{document}